\documentclass[%
 preprint,
 amsmath,amssymb,
 aps,
]{revtex4-1}

\usepackage{graphicx,psfrag}
\usepackage{caption} 
\usepackage{subcaption}
\usepackage{natbib}
\usepackage{float}
\usepackage{rotating}
\usepackage{mathrsfs,amsmath}
\usepackage{upgreek}
\usepackage{latexsym}
\usepackage{mathrsfs}
\usepackage{layout}
\usepackage{bm}
\usepackage{url}
\usepackage{wrapfig}
\usepackage{tabularx}
\usepackage{fancyhdr}
\usepackage{color}
\usepackage[normalem]{ulem}
\usepackage[section]{placeins}

\providecommand\bcdot{\boldsymbol{\cdot}}

\graphicspath{{./figures/}}

\begin{document}
	
	
	\title[]{Enhanced and suppressed multiscale dispersion of bidisperse inertial particles due to gravity}
	
	\author{Rohit Dhariwal}
	\author{Andrew D. Bragg}
	\email{andrew.bragg@duke.edu}
	\affiliation{Department of Civil and Environmental Engineering, Duke University, Durham, North Carolina 27708, USA}	
	\date{\today}

	\begin{abstract}
		
Using Direct Numerical Simulations (DNS), we investigate how gravity modifies the multiscale dispersion of bidisperse inertial particles in isotropic turbulence. The DNS has a Taylor Reynolds number $R_\lambda=398$, and we simulate Stokes numbers (based on the Kolmogorov timescale) in the range $St\leq 3$ , and consider Froude numbers $Fr = 0.052$ and $\infty$, corresponding to strong gravity and no gravity, respectively. The degree of bidispersity is quantified by the difference in the Stokes number of the particles $|\Delta St|$. We first consider the mean-square separation of bidisperse particle-pairs and find that without gravity (i.e. $Fr = \infty$), bidispersity leads to an enhancement of the the mean-square separation over a significant range of scales. When $|\Delta St|\geq O(1)$, the relative dispersion is further enhanced by gravity due to the large difference in the settling velocities of the two particles. However, when $|\Delta St|\ll1$, gravity suppresses the relative dispersion as the settling velocity contribution is small, and gravity suppresses the non-local contribution to the particle dynamics. In order to gain further insights, we consider separately the relative dispersion in the vertical (parallel to gravity) and horizontal directions. As expected, the vertical relative dispersion can be strongly enhanced by gravity due to differences in the settling velocities of the two particles. However, a key finding of our study is that gravity can also significantly enhance the horizontal relative dispersion. This non-trivial effect occurs because fast settling particles experience rapid fluctuations in the fluid velocity field along their trajectory, leading to enhanced particle accelerations and relative velocities. For sufficiently large initial particle separations, however, gravity can lead to a suppression of the horizontal relative dispersion. We also compute the Probability Density Function (PDF) of the particle-pair dispersion. Our results for these PDFs show that even when $Fr\ll1$ and $|\Delta St|\geq O(1)$, the vertical relative dispersion of the particles can be strongly affected by turbulence. This occurs because although the settling velocity contribution to the relative motion is much larger than the ``typical'' velocities of the turbulence when $Fr\ll1$ and $|\Delta St|\geq O(1)$, due to intermittency, there are significant regions of the flow where the turbulent velocities are of the same order as the settling velocity. These findings imply that in many applications where $R_\lambda\ggg1$, the effect of turbulence on the vertical relative dispersion of settling bidisperse particles may never be ignored, even if the particles are settling rapidly.
		
	\end{abstract}
	
	\maketitle

\section{Introduction}\label{sec:intro}
Particle dispersion in turbulent flows is important for numerous industrial and environmental applications such as drug delivery \citep{li96}, spray combustion \citep{faeth96}, plankton distribution in aquatic environments \citep{delillo14}, dispersion of pollutants in the atmosphere \citep{csanady} and droplet growth in warm clouds \citep{devenish12}. The relative dispersion of particles as a function of time is of particular importance as it provides a way to quantify multiscale processes such as particle mixing \citep{sawford05}, and it is also intimately connected to understanding particle collision velocities in turbulence \citep[e.g., see][]{pan10}.

Since the pioneering studies of \cite{richardson26} and \cite{batchelor52b}, the relative dispersion of fluid particles in turbulence has been extensively investigated. Nevertheless, many open questions remain \citep{salazar09}, and it continues to be a very active area of numerical, experimental and theoretical research \citep[e.g.][]{sawford05,biferale05,berg06a,salazar09,buaria15,bragg16,bragg17b,bragg18,dhariwal18prf,sawford08,ouellette06c,falkovich01,biferale14}.

In many applications, the dispersing particles have non-negligible inertia. The effect of this inertia on the relative dispersion of the particles in turbulence has only recently started receiving attention \citep{bec10b,biferale14,bragg16,bragg17b,bragg18}, and these studies have revealed that particle inertia can have striking effects on the relative dispersion. However, these studies focussed on monodisperse particles in the absence of gravity, yet in most real systems, the particles are polydisperse and are also settling under the effect of gravity. Since relative dispersion is typically studied by analyzing two-particle motion \citep{salazar09}, then in order to obtain a better understanding of more realistic systems, the relative dispersion of settling bidisperse particle-pairs should be considered. 

The importance of gravity compared with turbulence for particle motion can be quantified by the Froude number, $Fr \equiv \epsilon^{3/4}/(\nu^{1/4}g)$, where $\epsilon$ is the mean fluid kinetic energy dissipation rate, $\nu$ is the kinematic viscosity, and $g$ is the magnitude of the gravitational acceleration vector $\bm{g}$. In atmospheric clouds, typical ranges are $0.05\leq Fr\leq 0.3$  \citep{siebert10,ireland16b}. In a recent study \citep{dhariwal18jfm}, we considered the relative motion of settling, bidisperse particles in isotropic turbulence. In agreement with other studies \citep{zww01,chen17,kruis97}, we found that bidispersity alone enhances the particle relative velocities and these relative velocities are further enhanced by gravity. Furthermore, we also found that gravity can enhance relative velocities not only in the `vertical' (in the direction of gravity), but in the `horizontal' (in the plane normal to gravity) directions as well. Our results also showed that when $Fr \ll 1$, turbulence plays an important role, not only on the horizontal motion, but also on the vertical motion of particles since due to intermittency, there a significant regions of the flow where the local fluid acceleration is $O(g)$. These findings could have significant implications for understanding the horizontal and vertical relative dispersion of settling, bidisperse particles in turbulence.

Despite the significant practical importance, the relative dispersion of settling, bidisperse inertial particles in turbulence has scarcely been addressed. The only study that we are aware of attempting to address this problem is \cite{chang15}. The results from their Direct Numerical Simulations (DNS) showed that for particles having initial separations in the dissipation range, bidisperse particles separate faster in the presence of gravity due to the different settling velocities of the two particles. They also observed that bidisperse particles with and without gravity separate ballistically at short times, but that in the presence of gravity, the relative dispersion can follow a ballistic growth even beyond the short time regime. However, their study did not comprehensively explore the effects of varying the level of bidispersity, and also focused on weakly inertial particles. Further, they did not consider how gravity affects the dispersion in the vertical and horizontal directions separately. Therefore, motivated by our recent findings in \cite{dhariwal18jfm} and the current knowledge gaps, we consider the effect of gravity of the vertical and horizontal relative dispersion of inertial particles in turbulence with Stokes numbers up to $St=3$.

The organization of the paper is as follows. In \S\ref{theory}, we consider theoretically the effect of bidispersity and gravity on the relative dispersion of particle-pairs at the small-scales of turbulence. In \S\ref{CompD}, we explain the numerical methods and parameters for our simulations. In \S\ref{results}, we present the results of our simulations, exploring how bidispersity and $Fr$ impact the particle relative dispersion. Finally, in \S\ref{conc}, we draw conclusions and highlight open issues that remain to be explored.

\section{Theoretical Considerations}\label{theory}

In this paper we are considering the relative dispersion of settling, bidisperse inertial particles whose density is much greater than that of the fluid in which they are suspended (i.e. ``heavy particles'', $\rho_p/\rho \gg 1$, where $\rho_p$ is the particle density and $\rho$ is the fluid density). This is relevant to understanding various kinds of heavy particle dispersion in the atmosphere, and droplet mixing in clouds \citep{shaw03,grabowski13}. We consider the particle loading to be sufficiently small so that the feedback of particles on the flow can be ignored (i.e. the system is `one-way' coupled) and the particles are assumed to be small (i.e., $d/\eta \ll 1$, where $d$ is the particle diameter and $\eta$ is the Kolmogorov length scale). Under these conditions, each particle is treated as a point particle whose motion obeys a simplified version of the equation by Maxey \& Riley \cite{maxey83}
\begin{align}
\dot{\bm{x}}^p(t) & \equiv \bm{v}^p(t), \label{eq:part1}\\
\ddot{\bm{x}}^p(t) & \equiv \dot{\bm{v}}^p(t) = \frac{\bm{u}(\bm{x}^p(t),t)-\bm{v}^p(t)}{\tau_p} + \bm{g}, \label{eq:part2}
\end{align}
where $\bm{x}^p(t)$ and $\bm{v}^p(t)$ are the particle position and velocity, respectively, $\bm{u}(\bm{x}^p(t),t)$ is the fluid velocity at the particle position, $\tau_p \equiv \rho_p d^2/18 \mu$ is the particle response time ($\mu$ is the fluid dynamic viscosity) and 
$\bm{g}$ is the gravitational acceleration. In this study, we will assume that particles are subjected to linear drag force, which is a valid assumption for settling particles with $St\leq O(1)$\citep{ireland16b}, and this is the range we restrict attention to in this study.

To explore the relative motion of particles in turbulence we consider the motion of a ``satellite'' particle relative to a ``primary'' particle, where the kinematic equation governing their separation vector $\bm{r}^p(t)$ is
\begin{equation}
\dot{\bm{r}}^p(t) \equiv {\bm{w}}^p(t),
\label{eq:eom_r} 
\end{equation}
with solution 
\begin{equation}
\bm{r}^p(t) = \bm{r}^p(0) + \int_0^t \bm{w}^p(s)\,ds,
\label{eq:r_sol}
\end{equation}
where $\bm{w}^p(t)$ is the particle-pair relative velocity. The relative dispersion of the particle pair may be quantified through the statistical evolution of $\bm{r}^p(t)$, which depends upon the behavior of $\bm{w}^p(t)$. In order to consider how bidispersity and gravity affect the relative dispersion of inertial particles in turbulence, we now summarize our recent study \citep{dhariwal18jfm} concerning the effects of bidispersity and gravity on $\bm{w}^p(t)$.
\subsection{Relative velocities}
The equation of relative motion can be obtained by subtracting \eqref{eq:part2} for the primary particle from that for the satellite particle. In \cite{dhariwal18jfm}, we derived the following result for $\bm{w}^p(t)$ in non-dimensional form (non-dimensionalized with Kolmogorov scale quantities) and assuming $t\gg St_2$
\begin{equation}
\bm{w}^p(t) =\frac{1}{St_2} \int_0^t {\rm e}^{-(t-s)/St_2}\Delta \bm{u}^p(s) \,ds -\Delta St Fr^{-1}\bm{e}_{g} +\frac{\Delta St}{St_2} \int_0^t {\rm e}^{-(t-s)/St_2}\bm{a}^{p}(s)\,ds,
\label{eq:sol_w}
\end{equation}
where $\Delta \bm{u}^p(s)$ is the difference in the fluid velocity at the two particle positions, $St_2\equiv\tau_p/\tau_\eta$ is the Stokes number of the satellite particle based on the Kolmogorov time scale $\tau_\eta$, $\Delta St \equiv St_1-St_2$, where $St_1$ is the Stokes number of the primary particle, $\bm{e}_{g}$ is the unit vector in the direction of gravity, and $\bm{a}^{p}(s)$ is the primary particle acceleration. 

In the monodisperse case, $\Delta St=0$, only the first term on the rhs of (\ref{eq:sol_w}) survives. This term depends upon the particle-pair separation $\bm{r}^p$ through $\Delta\bm{u}^p$, and through the integral $\bm{w}^p(t)$ depends upon $\Delta\bm{u}^p$ along the path-history of the particle-pair over the time-span $t-s\leq O(\tau_p)$. Since the statistics of $\Delta\bm{u}^p$ depend upon scale then this leads to a non-local effect whereby the statistics of $\bm{w}^p(t)$ at a given scale are affected not only by the characteristics of $\Delta\bm{u}$ at that scale, but also by the characteristics of $\Delta\bm{u}$ at larger scales, and this allows for $\|\bm{w}^p\|>\|\Delta\bm{u}\|$ (statisticslly). For $St\geq O(1)$, this path-history/non-local effect dominates the particle relative velocities in the dissipation range and gives rise to the formation of ``caustics'' \cite{wilkinson05}. Its effect weakens at larger scales, and precisely vanishes at scales larger than the integral length scale of the flow since at these scales the statistics of $\Delta\bm{u}$ are independent of separation. For monodisperse particles, gravity only affects $\bm{w}^p(t)$ implicitly through its effect on $\Delta\bm{u}^p$, as it modifies how the particles interact with the turbulent flow. Gravity reduces the correlation timescale of $\Delta\bm{u}^p$ as it causes the particles to fall through the flow, and as a result, it reduces the path-history effect (by shrinking the temporal correlation radius over which the particles are affected by their past interaction with the flow), resulting in a reduction of the relative velocities for monodisperse particles \citep{ireland16b}.

For bidisperse particles without gravity, i.e. $|\Delta St|>0$ and $Fr=\infty$, bidispersity  affects $\bm{w}^p(t)$ explicitly through the third integral (``acceleration term'') and implicitly through the first integral on the rhs of \eqref{eq:sol_w}. The particle relative velocity at a given separation will depend upon the competition between the separation dependent first integral and separation independent third integral, and as $\Delta \bm{u}^p$ decreases, on average, with decreasing separation, there will be a scale below which the third integral will be greater than the first one. This acceleration contribution will lead to relative velocities of bidisperse particles that exceed those of monodisperse particles at small-separations.

To state the effects of gravity on the relative velocity of bidisperse particles, we must first introduce some notation. We define the gravitational force to act in the $x_3$ direction, so that $\bm{e}_{g}=(0,0,1)$. This will be referred to as the ``vertical'' direction, whereas, $x_1$ and $x_2$ will be referred to as the ``horizontal'' directions. (Since we are considering isotropic turbulence, the statistics of the particle motion are axisymmetric about $\bm{e}_{g}$ when $Fr<\infty$). When $|\Delta St|>0$ and $Fr<\infty$, the effect of gravity on the first term on the rhs of (\ref{eq:sol_w}) is qualitatively the same as in the monodisperse case, described earlier. The second term on the rhs of (\ref{eq:sol_w}) describes the explicit effect of gravity and it increases as $Fr$ decreases and/or $|\Delta St|$ increases. This term, however, only acts in the vertical direction, implying that gravity only plays an implicit role in the horizontal directions. Third term involving the primary particle acceleration is implicitly affected by gravity. In \cite{ireland16b}, we showed that gravity can enhance $\bm{a}^p$, since settling particles experience fluid velocities along their trajectories that fluctuate more rapidly than they would if they were not settling, resulting in larger particle accelerations. The enhancements of the accelerations in the vertical direction were found to be smaller than those in the horizontal directions due to differences in the longitudinal and transverse integral lengthscales of the flow \citep{ireland16b}.

Finally, consider the limit $Fr\to 0$, for which the relative velocities in the vertical direction are deterministic and given by 
\begin{equation}
\begin{split}
{w}_3^p(t) =  -\Delta St/Fr,
\end{split}
\label{w3Fr0}
\end{equation}
which is simply the differential settling velocity, in dimensionless form, and applies to quiescent and turbulent flows. However, for the horizontal direction, in a quiescent flow $w^p_1(t)=0\,\forall Fr$, but in a turbulent flow, $w^p_1(t)\neq 0$ in the limit $Fr\to 0$ and is given by
\begin{equation}
\begin{split}
{w}_1^p(t) = &\frac{1}{St_2} \int_0^t {\rm e}^{-(t-s)/St_2}\Delta {u}_1^p(s) \,ds + \frac{\Delta St}{St_2} \int_0^t {\rm e}^{-(t-s)/St_2} {a}_1^p(s)\,ds.
\end{split}
\label{w1Fr0}
\end{equation}
As a consequence of this, $\lim_{Fr\to0}\bm{w}^p(t)\not\to-\Delta St Fr^{-1}\bm{e}_{g} $; turbulence always makes a contribution to the horizontal motion and this has important implications.

\subsection{Relative dispersion}\label{RD}

We now consider the implications of these results on $\bm{w}^p(t)$ for the relative dispersion of inertial particles in turbulence, focusing attention on the case where the initial separation $\bm{r}^p(0)=\bm{\xi}$ lies in the dissipation range, where the effects of inertia are the strongest. 

For monodisperse particles with $St_2\geq O(1)$ and $Fr=\infty$, the particles will initially separate (on average) very fast due to caustics in the particle relative velocity distributions. As $t$ increases, the particles go to larger separations where the effects of particle inertia become weaker, and eventually they will separate like fluid particles. This behavior was observed in \cite{bec10b,bragg16}. When $St_2\geq O(1)$ and $Fr\leq O(1)$, we expect that monodisperse inertial particles will separate slower than the fluid particles because gravity suppresses the path-history effect, and the dominant effect of inertia will be to simply filter out the fluctuations of the underlying flow.

For bidisperse particles ($|\Delta St|>0$) with  $Fr=\infty$, the acceleration term dominates their relative velocities at small separations and they will separate faster than the monodisperse particles. As $t$ increases, the particle-pairs will move to scales where the first term on the right hand side of \eqref{eq:sol_w} dominates their relative velocity, and their dispersion behavior will approach that of monodisperse particles. At sufficiently long times, the effect of their inertia will disappear and they will disperse as fluid particles. The strongest effects of bidispersity will therefore occur in the regime where the acceleration term in \eqref{eq:sol_w} makes a strong contribution to the bidisperse particle-pair motion, and the duration of time for which this term will be important will depend upon their initial separation and $|\Delta St|$. For example, for pairs with $|\Delta St|\gg 1$ and initial separation in the dissipation range, the acceleration term will play a dominant role in the relative dispersion process up until times for which the pair separation is well into the inertial range. 

When $Fr\leq O(1)$, gravity can affect the relative dispersion of bidisperse particles in both the horizontal and vertical directions. Clearly, the vertical dispersion will be affected explicitly by gravity due to the contribution of the differential settling velocity. However, the relative dispersion in the horizontal direction can also be strongly affected due to the implicit effect of gravity on $\bm{a}^p$. As summarized earlier, \cite{ireland16b} found that when $Fr \ll 1$, gravity can significantly enhance both the vertical and horizontal components of $\bm{a}^p$. Therefore, for bidisperse particles with small initial separations, gravity can enhance both the vertical and horizontal relative dispersion, in contrast to the monodisperse case, where gravity is expected to always lead to a suppression of the relative dispersion, in both the vertical and horizontal directions. 

However, for weakly bidisperse particles with $|\Delta St|\ll 1$, and/or for bidisperse particles with sufficiently large initial separations, the acceleration contribution to their relative velocities will be sub-dominant, and as a result, in these regimes, gravity may also lead to a suppression of the relative dispersion, especially in the horizontal direction (even in these regimes, the vertical relative dispersion may still be enhanced by gravity due to the contribution from the differential settling velocity).

Chang \emph{et al.} \cite{chang15} used DNS to study the dispersion of bidisperse particles with and without gravity. In their study, they considered pairs with initial separations in the dissipation range and $St_2\ll1$, $|\Delta St|\ll 1$ and $Fr = 0.1,\infty$. They found that the differential settling velocities of the particles leads to faster relative dispersion of bidisperse particles as compared to those dispersing in the absence of gravity. They also observed that bidisperse particles with and without gravity separate ballistically at short times, but that in the presence of gravity, the relative dispersion can follow a ballistic growth even beyond the short time regime. 
They also developed a semi-empricial formula for the second-order structure function $\langle\|\bm{w}^p(t)\|^2\rangle_{\bm{r}}$, where  $\langle\cdot\rangle_{\bm{r}}$ denotes an ensemble average conditioned on $\bm{r}^p(t)=\bm{r}$. By comparing their results with DNS data for $Fr=0.1$, they found good agreement between their prediction and the DNS data even when $St_2=O(1)$ (with $|\Delta St|\ll 1$). However, this is mainly because in their data with $Fr=0.1$,  $\langle\|\bm{w}^p(t)\|^2\rangle_{\bm{r}}\approx(\Delta St)^2 Fr^{-2}$ for small $r/\eta$, i.e. the turbulent contribution is negligible, and this gravitational settling contribution is in exact, closed form for arbitrary $St_2$, $\Delta St$.

\section{Computational Details}\label{CompD}
We perform DNS of statistically stationary, isotropic turbulence using a pseudospectral method on a three-dimensional periodic cubic domain of  length $\mathscr{L}$. The computational domain is uniformly discretized using $N^3$ grid points and the fluid velocity field $\bm{u}(\bm{x},t)$ is obtained by solving the incompressible Navier-Stokes equation
\begin{eqnarray}
\partial_t\bm{u} + {\bm{\omega}} \times \bm{u} 
+ \bm{\nabla}\left ( \frac{p}{\rho_f} + \frac{\|\bm{u}\|^{2}}{2}  \right )  = \nu \bm{\nabla}^2 \bm{u} + \bm{f}, \label{eq:ns}
\end{eqnarray}
where $\bm{\omega} \equiv \bm{\nabla} \times \mathbf{u}$ is the vorticity, $\rho_f$ is the fluid
density, $p$ is the pressure (determined using $\bm{\nabla} \bcdot \bm{u} = 0$), $\nu$ is the kinematic viscosity and  $\bm{f}$ is the external forcing applied to generate statistically stationary turbulence.
A deterministic forcing scheme is used for $\bm{f}$ \citep {witkowska97}, where the energy dissipated during one time step is resupplied to the 
wavenumbers with magnitude $\kappa \in (0, \sqrt{2}]$. Time integration is performed through a second-order, explicit Runge-Kutta scheme with aliasing errors removed by means of a combination of 
spherical truncation and phase-shifting. 
\begin{table}
\centering
\renewcommand{\arraystretch}{1.1}
\setlength{\tabcolsep}{32pt}
  \begin{tabular}{cc}
  \hline
    $\mathrm{Parameter}$ & $\mathrm{DNS}$ \\
       $N$ & 1024 \\
       $R_\lambda$ &  398 \\
       $\mathscr{L}$ & 2$\pi$ \\       
       $\nu$ & 0.0003 \\
       $\epsilon$ & 0.223 \\
       $l$ & 1.45 \\
       $l/\eta$ & 436 \\
       $u'$ & 0.915 \\
       $u'/u_\eta$ & 10.1 \\
       $T_L$ & 1.58 \\
       $T_L/\tau_\eta$ & 43.0 \\
       $\kappa_{{\rm max}}\eta$ & 1.60 \\
       $N_p$ & 2,097,152\\
       \hline
  \end{tabular}
  \caption{Flow parameters in DNS of isotropic turbulence (arbitrary units).
          $N$ is the number of grid points in each direction, 
          $R_\lambda \equiv u'\lambda/\nu$ is the Taylor micro-scale
          Reynolds number,
          $\mathscr{L} $ is the domain size, $\nu$ is the fluid kinematic viscosity, $\epsilon \equiv 2\nu \int_0^{\kappa_{\rm max}}\kappa^2 E(\kappa) {\rm d}\kappa $ is the mean
          turbulent kinetic energy dissipation rate,
          $l \equiv 3\pi/(2k)\int_0^{\kappa_{\rm max}}E(\kappa)/\kappa {\rm d}\kappa $  is the integral length scale, $\eta \equiv \nu^{3/4}/\epsilon^{1/4}$ is the Kolmogorov length scale, 
           $u' \equiv \sqrt{(2k/3)}$ is the fluid r.m.s. fluctuating 
          velocity, $k$ is the turbulent kinetic energy, 
          $u_\eta$ is the Kolmogorov velocity scale, 
          $T_L \equiv l/u^\prime$ is the large-eddy turnover
          time, $\tau_\eta \equiv \sqrt{(\nu/\epsilon)}$ is the Kolmogorov time scale, 
          $\kappa_{\rm max}$ is the maximum
          resolved wavenumber, and $N_p$ is  
          the number of particles per Stokes number.}
  {\label{tab:parameters}}
  
\end{table}

Particles are tracked in the flow field using \eqref{eq:part2} for their equation of motion, using a ``one-way'' coupled assumption, which is justified in the cloud context, for example, because of the low particle loading. Fifteen particle classes are simulated with Stokes number $St$ ranging from 0 to 3, with $(N/8)^3$ particles per $St$. The particles are introduced to the domain once the underlying fluid has achieved statistically stationary state, with initial particle velocities equal to the fluid velocity at the particle position. The particles are allowed to evolve for nearly five large eddy turnover times before we start collecting statistics. The particle positions and velocities are stored every $0.1\tau_\eta$ for a duration of $100\tau_\eta$.  

In \eqref{eq:part2}, $\bm{u}(\bm{x}^p(t),t)$ is the fluid velocity at the particle position, and this must be evaluated by interpolating the fluid velocity at the surrounding grid points
to $\bm{x}^p(t)$. In this study we use an $8^{th}$-order, B-spline interpolation method which provides a good balance between high-accuracy and efficiency (see \cite{ireland13}). Further details on all aspects of the computational methods can be found in \cite{ireland13}. 

Since we want to explore the role of gravity on the relative dispersion of bidisperse particles, we must consider the choice of $Fr$ for the DNS. Observations have shown that $\epsilon$ can vary by orders of magnitude in clouds \citep{prupp97}, with corresponding variations in $Fr$. Therefore, in addition to the zero gravity case $Fr=\infty$, we follow \cite{ireland16b} and consider $Fr= 0.052$, which may be considered to be representative of weakly turbulent stratiform clouds \citep{pinsky07}. 

In our recent studies \cite{ireland16b,dhariwal18jfm} we highlighted that the use of periodic boundary conditions in the DNS, while simulating very small values of $Fr$, can artificially influence the motion of inertial particles if the DNS box length $\mathscr{L}$ is too small. The use of periodic boundary conditions is problematic if the time it takes the settling particles to traverse the distance $\mathscr{L}$ is $\leq O(T_L)$, where $T_L$ is the large eddy turnover time. This issue was explored in detail in \cite{ireland16b} and it was found that box sizes much larger than the standard $\mathscr{L}=2\pi$ can be required when $Fr$ is very small. For example, in \cite{ireland16b} it was found that for $R_\lambda\approx 90$, $\mathscr{L}=16\pi$ was necessary, which, with the resolution constraints for accurately resolving the small-scales requires $N=1024$. Such requirements place significant limitations on the value of $R_\lambda$ that can be simulated. The box size issue is even more crucial when considering relative dispersion since the relative dispersion can only be tracked for times for which the pair-separation is smaller than the box size, and this may not be very long when $|\Delta St|/Fr\gg 1$.

In this study we consider $R_\lambda\approx 398$, and for $Fr\geq 0.052$ and $St\leq 3$, a box size of length $\mathscr{L}=2\pi$ is sufficient to satisfy the aforementioned constraint on the use of periodic boundaries for settling particles (see \cite{ireland16b}). Details of the DNS are summarized in Table~\ref{tab:parameters}.

\section{Results and discussion}\label{results}
\FloatBarrier
In this section we discuss our DNS results for the relative dispersion of bidisperse inertial particles with and without gravity. We consider particle pairs ($St_1, St_2$ combination) with varying degree of bidispersity, namely, weak, moderate and strong bidispersity, corresponding to $|\Delta St|$ = 0.1, 0.5 and 2, respectively. With $Fr\ll1$ and for $|\Delta St|= 2$, the particle separation quickly approaches the size of the computational domain, beyond which the relative dispersion results are artificially influenced by the periodic boundary conditions. Consequently, for those cases, the results are only shown up to the times at which the separations are affected by the domain size.
\begin{figure}
	\centering
	\includegraphics[width=0.6\textwidth]{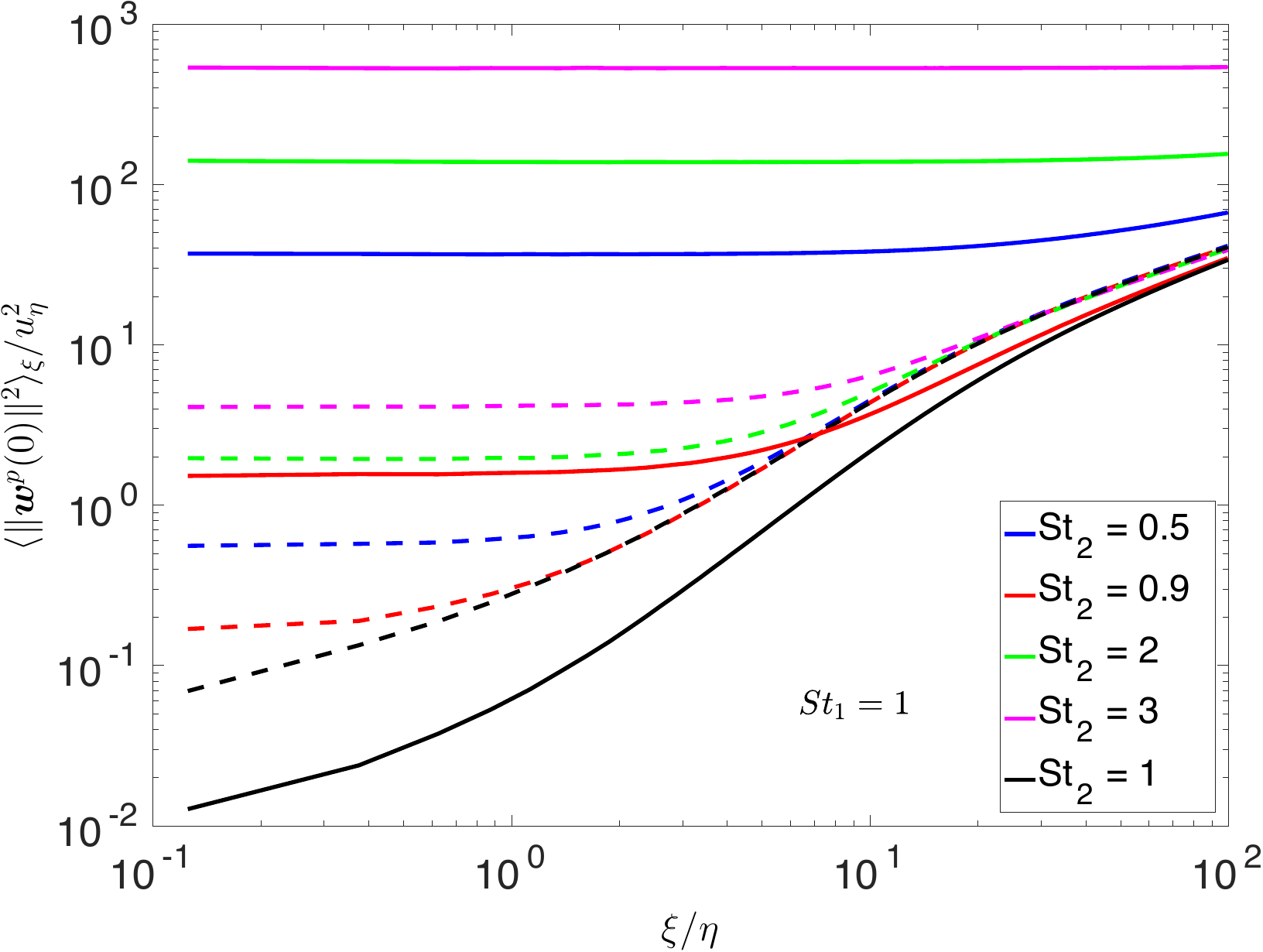}
		\caption{ $\langle \|\bm{w}^p({0})\|^2\rangle_{\xi}/u_\eta^2$ as a function of $\xi/\eta$ for $St_1 = 1$ and various $St_2$. Solid and dashed lines correspond to the results with $Fr = 0.052~\textrm{and}~\infty$, respectively. }
	\label{fig:w0}	
\end{figure}
%
\FloatBarrier
We begin by considering the DNS data for particle relative velocity $\langle \|\bm{w}^p({0})\|^2\rangle_{\xi}$ (where $\langle\cdot\rangle_{\xi}$ denotes an ensemble average conditioned on $\|\bm{r}^p(0)\|=\xi$), as this statistic  will be helpful in understanding the relative dispersion results, especially in the short-time regime. Figure \ref{fig:w0} shows the variation of  $\langle \|\bm{w}^p({0})\|^2\rangle_{\xi}$ as a function of $\xi$ for bidisperse particles, with $Fr = \infty$ and $Fr=0.052$. It can be seen that the bidisperse data, with and without gravity, are bounded from below by the monodisperse values. This observation is in agreement with previous numerical and theoretical studies considering the relative velocities of bidisperse particles in turbulence \citep{dhariwal18jfm,zww01,zaichik06a,zaichik09b,pan10,pan14}. It can also be noticed that at sufficiently small separations the relative velocities become independent of $\xi$. Both of these effects are due to the contribution of the acceleration term to the bidisperse particle motion, as discussed in \S\ref{theory}. 
The relative velocities of bidisperse particles at small scales increase with $|\Delta St|$ and are further enhanced by gravity (as $Fr$ is decreased). The effect of bidispersity reduces as the pair separation increases, and at sufficiently large $\xi$ the relative velocities of bidisperse particles approach those of monodisperse particles. This is because at large separations the first term on the right-hand side of \eqref{eq:sol_w} becomes dominant as compared to the third term (acceleration term).  
\begin{figure}
	\centering
	\includegraphics[width=\textwidth]{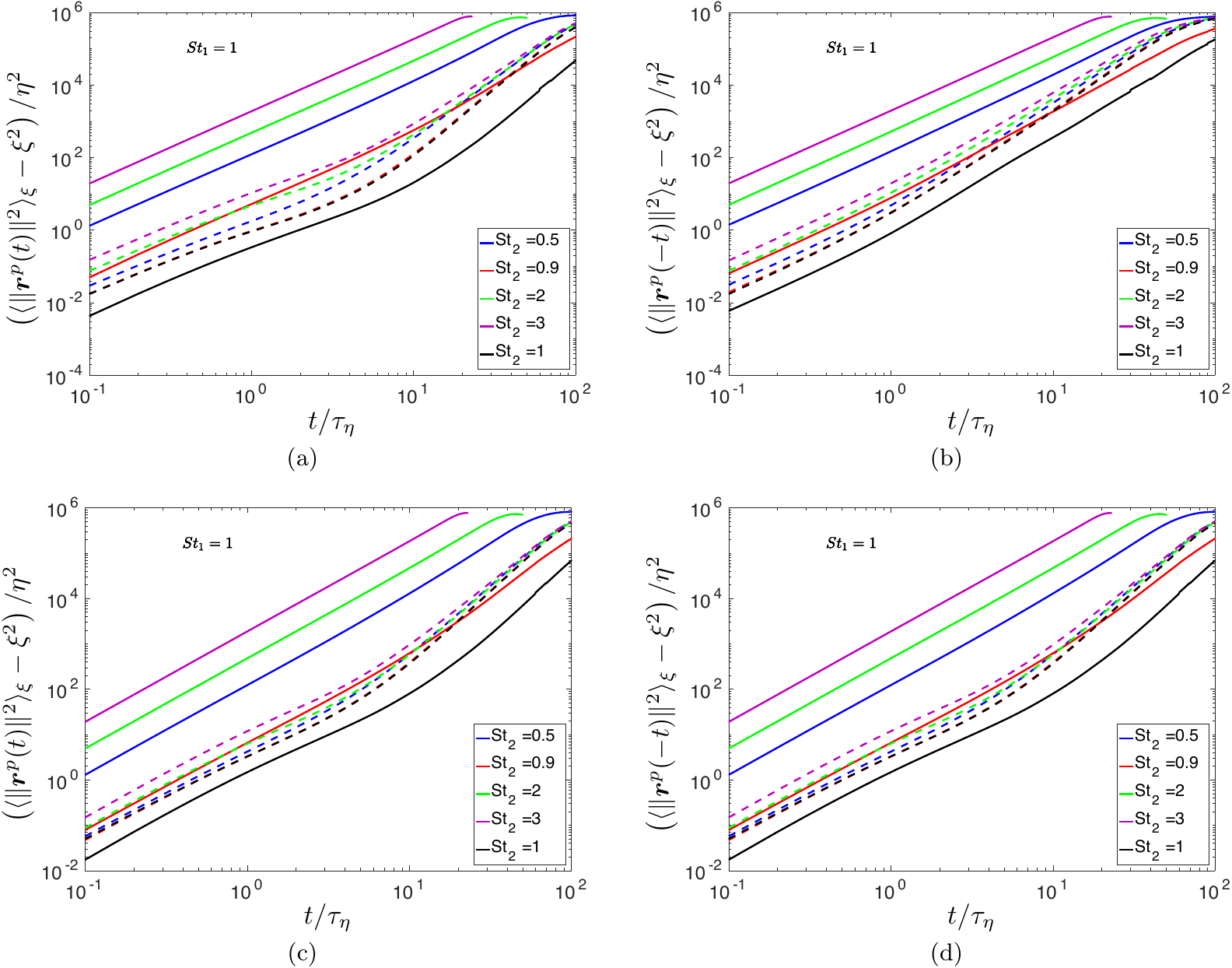}
		\caption{ FIT and BIT mean-square separation results (with $\xi^2$ subtracted) for $St_1 = 1$ and various $St_2$, (a) and (b) $\xi \in (0\eta,1\eta]$, (c) and (d) $\xi \in [3\eta,4\eta]$. Solid and dashed lines correspond to the results with $Fr = 0.052~\textrm{and}~\infty$, respectively.}
	\label{fig:mss_ssep}	
\end{figure}
	\FloatBarrier
We now turn to consider the mean-square separation results, for both Forward In Time (FIT) $\langle \|\bm{r}^p({t})\|^2\rangle_{\xi}$ and Backward In Time (BIT) $\langle \|\bm{r}^p({-t})\|^2\rangle_{\xi}$. The results are shown in figure \ref{fig:mss_ssep} for $\xi/\eta \in (0,1]$, $\xi/\eta \in [3,4]$, for both monodisperse and bidisperse particles. In general, the results show that both FIT and BIT relative dispersion are enhanced with increasing $|\Delta St|$, and are further enhanced by gravity. However, for monodisperse and weakly bidisperse particles, gravity leads to a suppression of the relative dispersion, as anticipated in our discussion in \S\ref{RD}. The results also show that FIT and BIT dispersion are qualitatively and quantitatively different, indicating irreversibility in the relative dispersion process. 
\begin{figure}
	\centering
	\includegraphics[width=\textwidth]{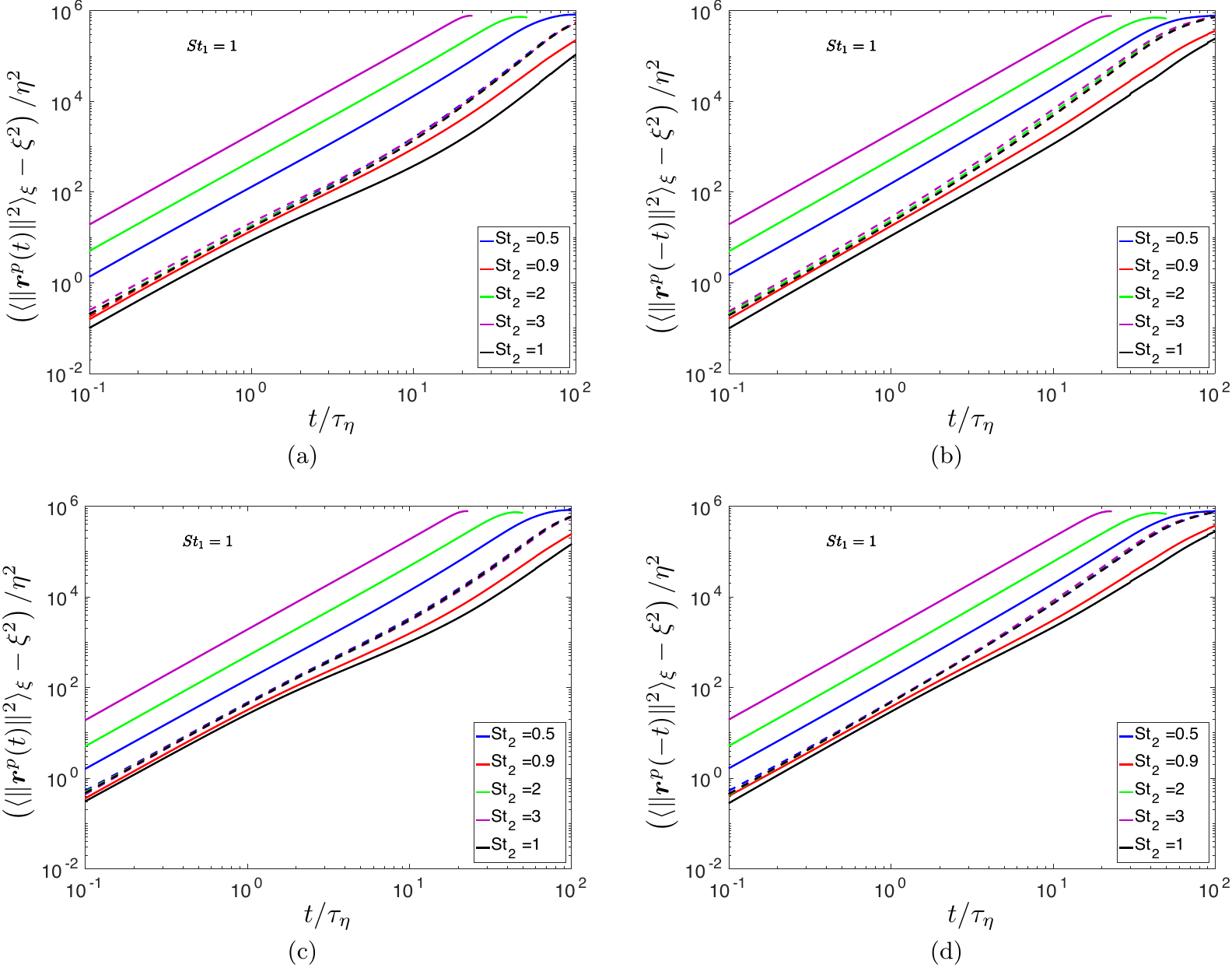}
		\caption{ FIT and BIT mean-square separation results (with $\xi^2$ subtracted) for $St_1 = 1$ and various $St_2$, (a) and (b) $\xi \in [9\eta,10\eta]$, (c) and (d) $\xi \in [19\eta,20\eta]$. Solid and dashed lines correspond to the results with $Fr = 0.052~\textrm{and}~\infty$, respectively.}
	\label{fig:mss_lsep}
\end{figure}
\FloatBarrier
The physical mechanisms responsible for this irreversibility are quite subtle, and we therefore refer the reader to \cite{bragg16,bragg17b,bragg18}. In figure \ref{fig:mss_lsep} we show results for larger $\xi$ and observe that without gravity, bidispersity has a weak effect on the dispersion at these separations. Also, for monodisperse and weakly bidisperse particles, gravity suppresses the relative dispersion at all times for these larger $\xi$ values.
\begin{figure}
	\centering
	\includegraphics[width=\textwidth]{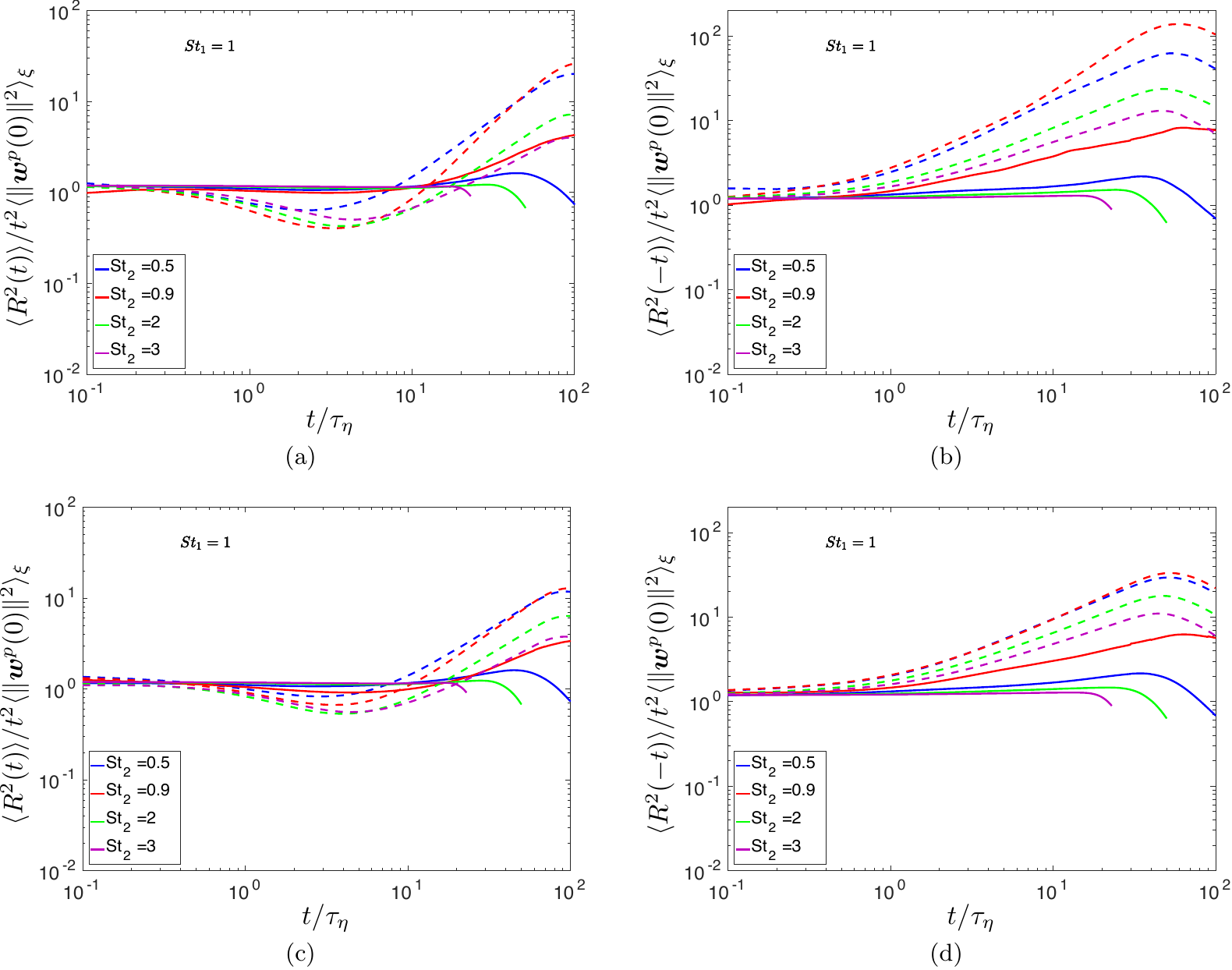}
		\caption{FIT and BIT results for $\langle {R}^2(t) \rangle \equiv \langle \|\bm{r}^p({t})\|^2\rangle_{\xi}-\xi^2$, divided by the ballistic prediction $t^2\langle \|\bm{w}^p({0})\|^2\rangle_{\xi}$ for  $St_1 = 1$ and various $St_2$ combinations, (a) and (b) $\xi \in (0\eta,1\eta]$, (c) and (d) $\xi \in [3\eta,4\eta]$. Solid and dashed lines correspond to the results with $Fr = 0.052~\textrm{and}~\infty$, respectively.}
	\label{fig:mss_ballistic}
\end{figure}
\FloatBarrier
In figure \ref{fig:mss_ballistic}, we plot $\langle \|\bm{r}^p({t})\|^2\rangle_{\xi}-\xi^2$ and $\langle \|\bm{r}^p({-t})\|^2\rangle_{\xi}-\xi^2$, normalized by the short-time ($t\ll\tau_p$) ballistic predictions \citep[e.g.,][]{bragg16}
\begin{align}
\langle \|\bm{r}^p({t})\|^2\rangle_{\xi}-\xi^2\approx \langle \|\bm{r}^p({-t})\|^2\rangle_{\xi}-\xi^2=t^2\langle \|\bm{w}^p({0})\|^2\rangle_{\xi}+O(t^3),\label{ballpredic}
\end{align}

The results in figures \ref{fig:mss_ballistic}(a) and (c) show that the mean-square separation grows ballistically for short times in the absence of gravity. However, for $Fr=0.052$, \eqref{ballpredic} can be satisfied even for relatively long times in the dispersion process. This is explained by the fact that in view of \eqref{eq:sol_w}, in the regime $\vert \Delta St\vert Fr^{-1}\gg 1$, the differential settling of the particle-pair dominates their relative dispersion and we have (in dimensional form)
\begin{align}
\langle \|\bm{r}^p({t})\|^2\rangle_{\xi}-\xi^2= \langle \|\bm{r}^p({-t})\|^2\rangle_{\xi}-\xi^2\approx \left(u_\eta\Delta St/Fr\right)^2 t^2,
\end{align}
which applies for as long as the differential sedimentation dominates $\bm{w}^p$.

\subsection{Horizontal and vertical separations}
%
\vspace{0mm}
\begin{figure}
	\centering
	\includegraphics[width=\textwidth]{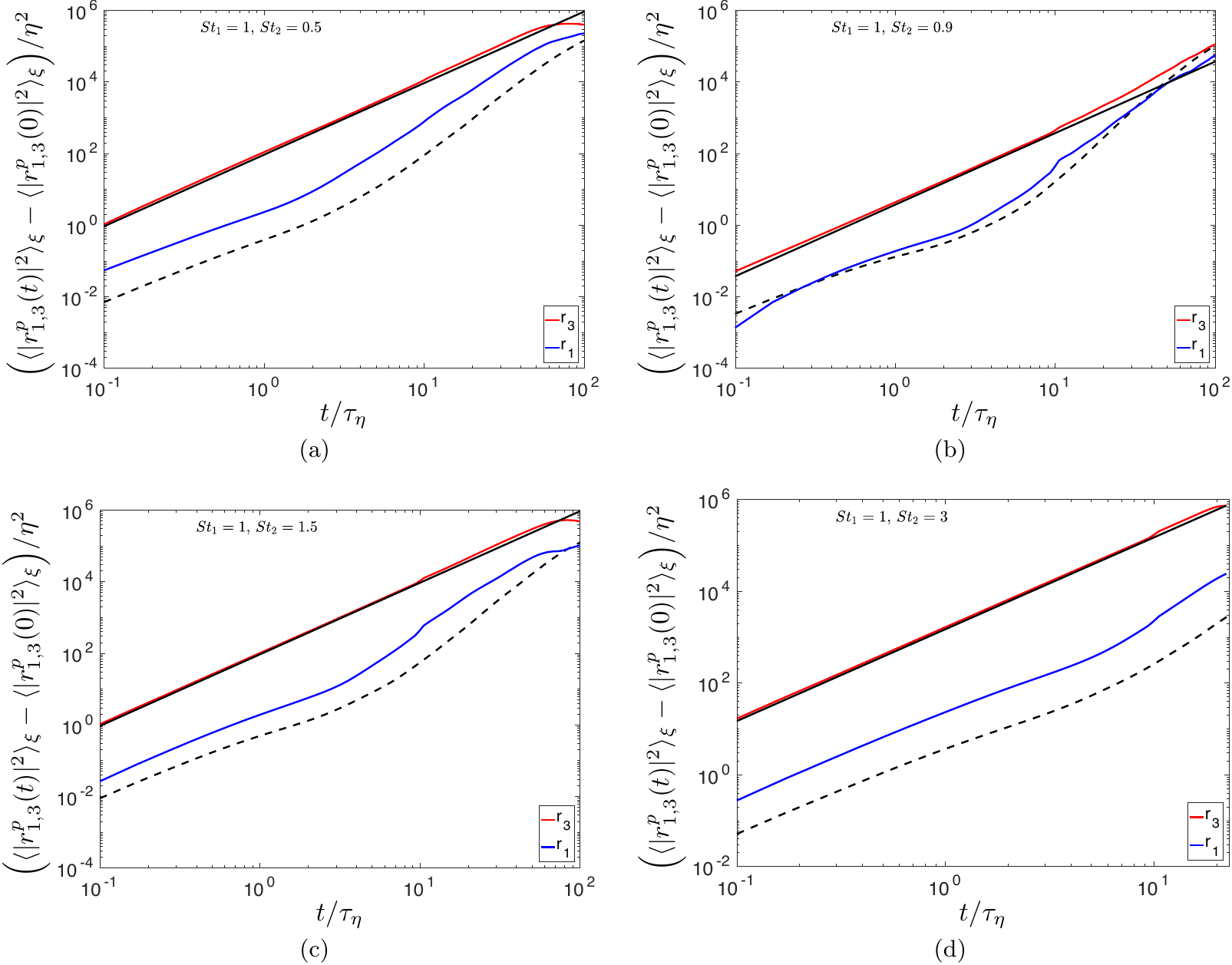}
		\caption{FIT mean-square separation results in the vertical and horizontal directions from DNS for $St_1 = 1$, different $St_2$, and with $\xi \in [0\eta,1\eta]$. Red line corresponds to the vertical separations and the blue line corresponds to the horizontal separations. Dashed black line corresponds to the results without gravity. The solid black line corresponds to \eqref{r3r3} for $Fr = 0.052$.}
	\label{fig:fit_r13_1eta}
\end{figure}
%
In order to untangle the explicit and implicit effects of gravity on the statistics of $\langle \|\bm{r}^p({t})\|^2\rangle_{\xi}$ and gain further insight, we consider the mean-square separations based on the Cartesian components of $\bm{r}^p({t})$. That is, we consider $\langle |r^p_3({t})|^2\rangle_{\xi}$ and $\langle |r^p_1({t})|^2\rangle_{\xi}$, corresponding to the separations in the vertical and horizontal directions, respectively (recall also that $\langle |r^p_1({t})|^2\rangle_{\xi}=\langle |r^p_2({t})|^2\rangle_{\xi}$ due to axisymmetry of the statistics).

\begin{figure}
	\centering
	\includegraphics[width=\textwidth]{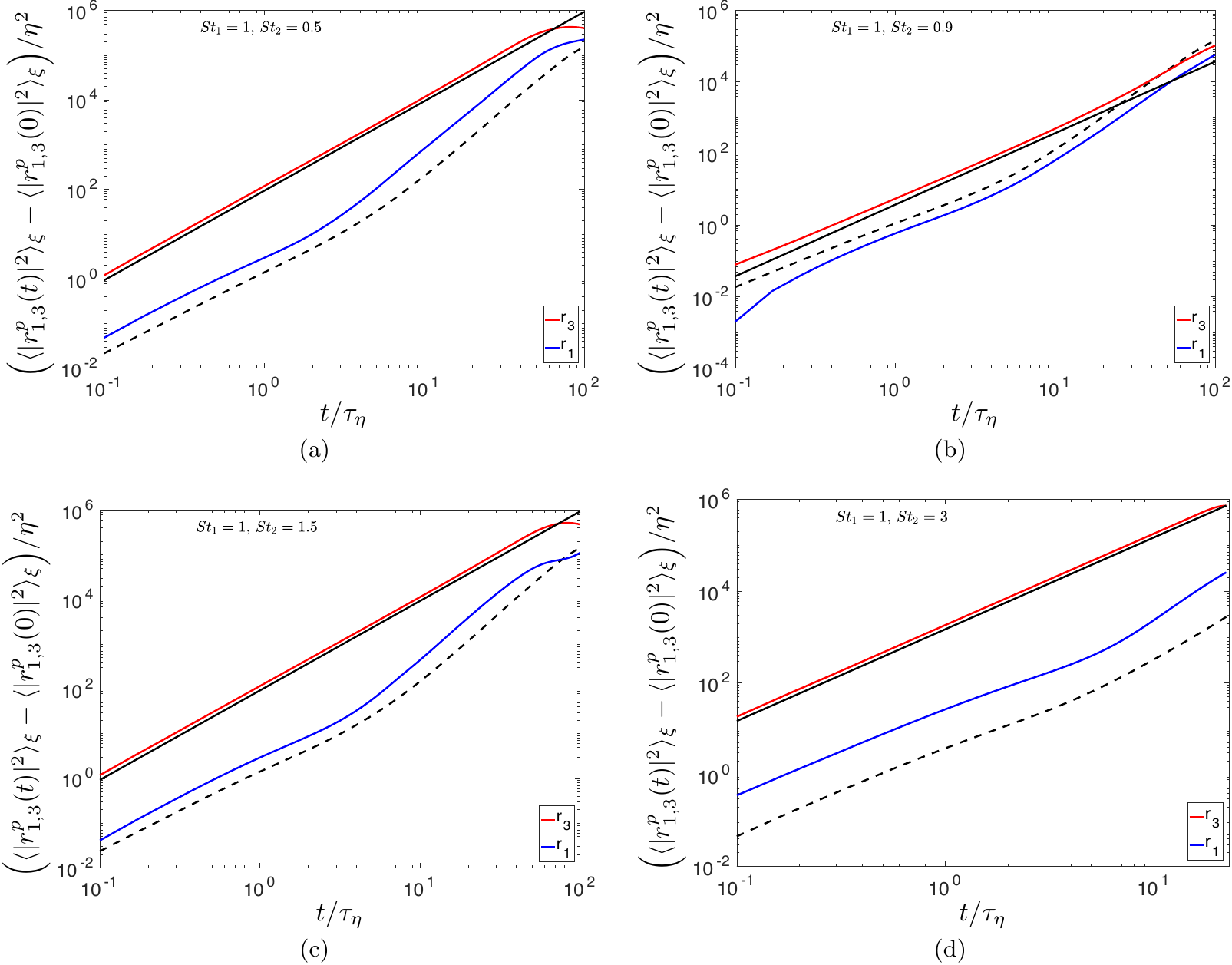}
		\caption{FIT mean-square separation results in the vertical and horizontal directions from DNS for $St_1 = 1$, different $St_2$, and with $\xi \in [3\eta,4\eta]$. Red line corresponds to the vertical separations and the blue line corresponds to the horizontal separations. Dashed black line corresponds to the results without gravity. The solid black line corresponds to \eqref{r3r3} for $Fr = 0.052$.}
	\label{fig:fit_r13_4eta}
	\end{figure}

The FIT results for $\xi \in [0\eta,1\eta]$ and $\xi\in[3\eta,4\eta]$ are shown in figures~\ref{fig:fit_r13_1eta} and \ref{fig:fit_r13_4eta}, where we also show the $Fr\to0$ prediction for the vertical dispersion
\begin{align}
\langle |r^p_3(t)|^2 \rangle_{\xi}-\langle |r^p_3(0)|^2 \rangle_{\xi}=\left(u_\eta\Delta St /Fr\right)^2 t^2.\label{r3r3}
\end{align}
For $Fr=0.052$, \eqref{r3r3} is in almost perfect agreement with the data for all the $St_1, St_2$ combinations having $|\Delta St| \geq 0.5$. The results show that both the horizontal and vertical dispersion are enhanced as $Fr$ is reduced. As discussed in \S\ref{theory}, while the enhancement of the vertical relative dispersion is mainly due to the differential sedimentation of the particles, the enhancement of the horizontal relative dispersion occurs only through the the implicit effect of gravity that enhances the particle accelerations, and through this the relative velocities. The results also show that the gravity-driven enhancement of ${\langle [r^p_1(t)]^2 \rangle_{\xi}}$ can persists up to long times in the dispersion process. Indeed, the enhancement will persist for as long as the particle-pair remains at scales where the acceleration contribution to ${w}_1^p$ continues to be significant. For weakly bidisperse particle-pairs ($|\Delta St| \ll1$), gravity suppresses ${\langle [r^p_1(t)]^2 \rangle_{\xi}}$ by suppressing $w^p_1$. As explained in \S\ref{theory}, this is because when $|\Delta St| \ll1$, ${w}_1^p$ is dominated by the path-integral involving $\Delta u^p_1$ rather than the acceleration term, and the statistics of this path-integral are reduced by gravity, since gravity reduces the correlation timescales of $\Delta u^p_1$ \citep{ireland16b,dhariwal18jfm}. We note however, that for sufficiently large $R_\lambda$, then irrespective of $|\Delta St|$, the particles will eventually disperse to scales that are large enough for the effects of bidispersity to be weak (i.e. the acceleration contribution to ${w}_1^p$ would be small), at which point the effect of gravity would be to suppress the relative dispersion.
 
\begin{figure}
	\centering
	\includegraphics[width=\textwidth]{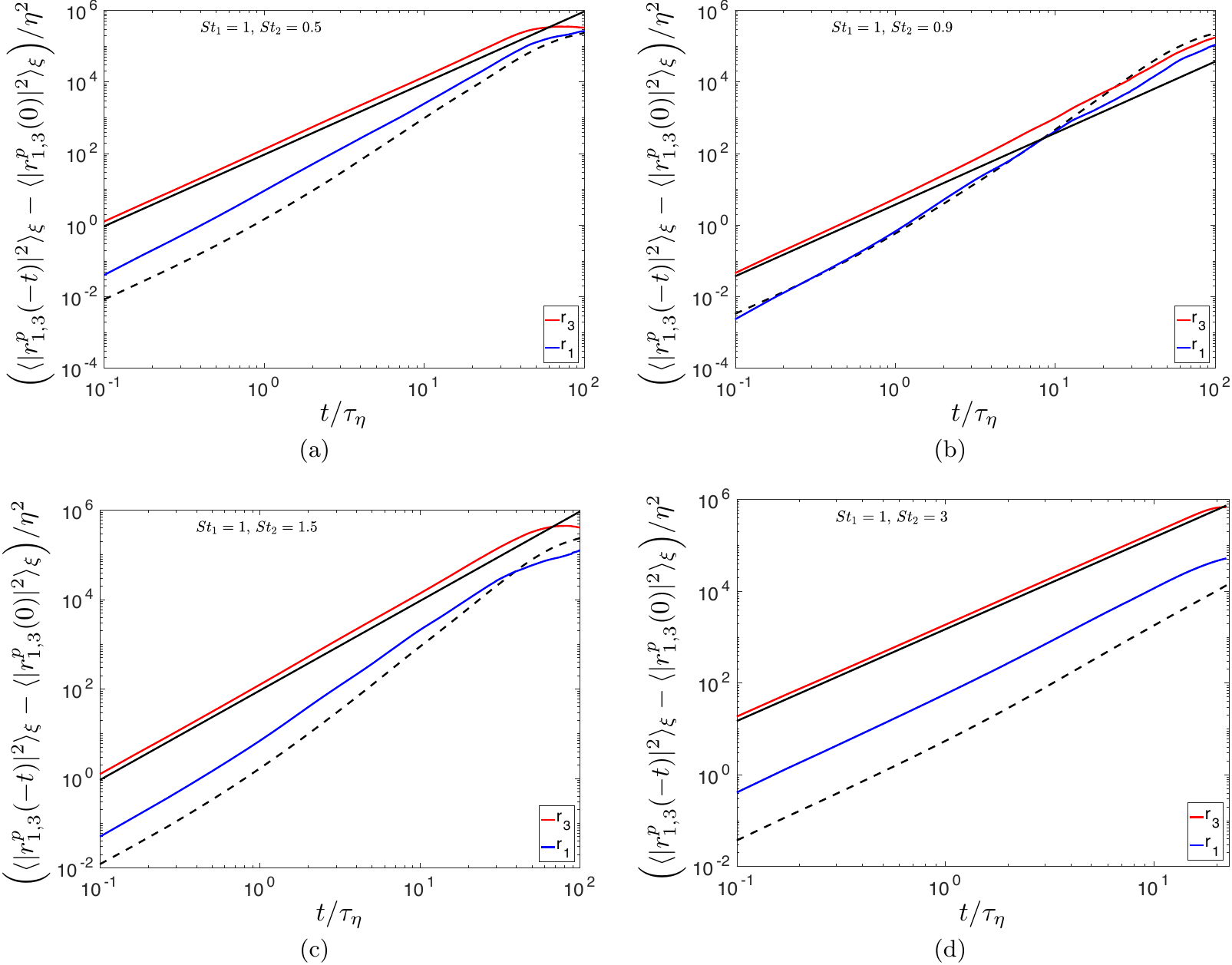}
		\caption{BIT mean-square separation results in the vertical and horizontal directions from DNS for $St_1 = 1$, different $St_2$, and with $\xi \in [0\eta,1\eta]$. Red line corresponds to the vertical separations and the blue line corresponds to the horizontal separations. Dashed black line corresponds to the results without gravity. The solid black line corresponds to \eqref{r3r3} for $Fr = 0.052$.}
	\label{fig:bit_r13_1eta}
	\end{figure}

\begin{figure}
	\centering
	\includegraphics[width=\textwidth]{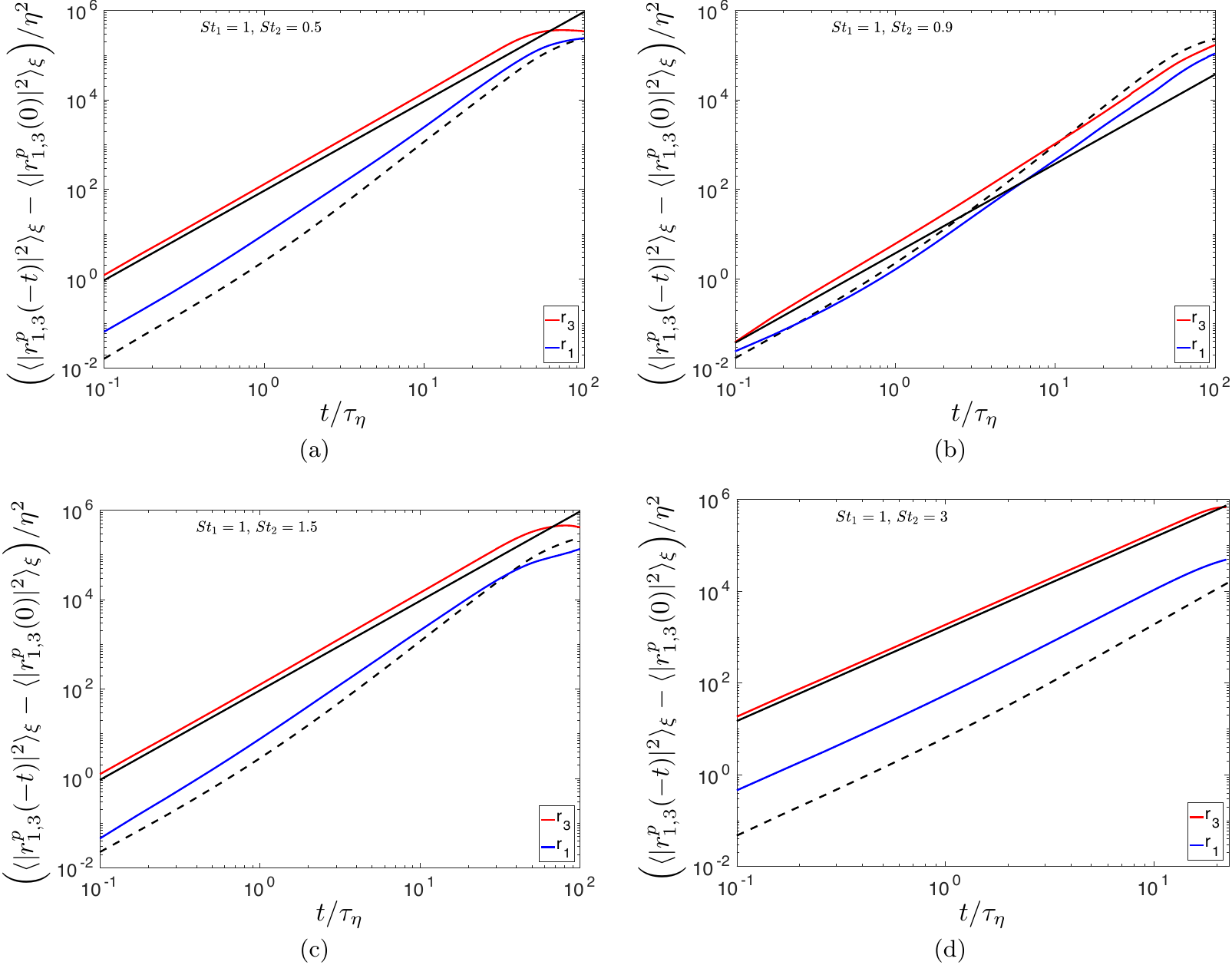}
		\caption{BIT mean-square separation results in the vertical and horizontal directions from DNS for $St_1 = 1$, different $St_2$, and with $\xi \in [3\eta,4\eta]$. Red line corresponds to the vertical separations and the blue line corresponds to the horizontal separations. Dashed black line corresponds to the results without gravity. The solid black line corresponds to \eqref{r3r3} for $Fr = 0.052$.}
	\label{fig:bit_r13_4eta}
	\end{figure}

The results for BIT horizontal and vertical dispersion are shown in figures \ref{fig:bit_r13_1eta} and \ref{fig:bit_r13_4eta}, and they also show that gravity has the same qualitative effect as in the FIT case, enhancing and suppressing the relative dispersion in different regimes.

\subsection{PDFs of horizontal and vertical pair separations}
\begin{figure}
	\centering
	\includegraphics[width=\textwidth]{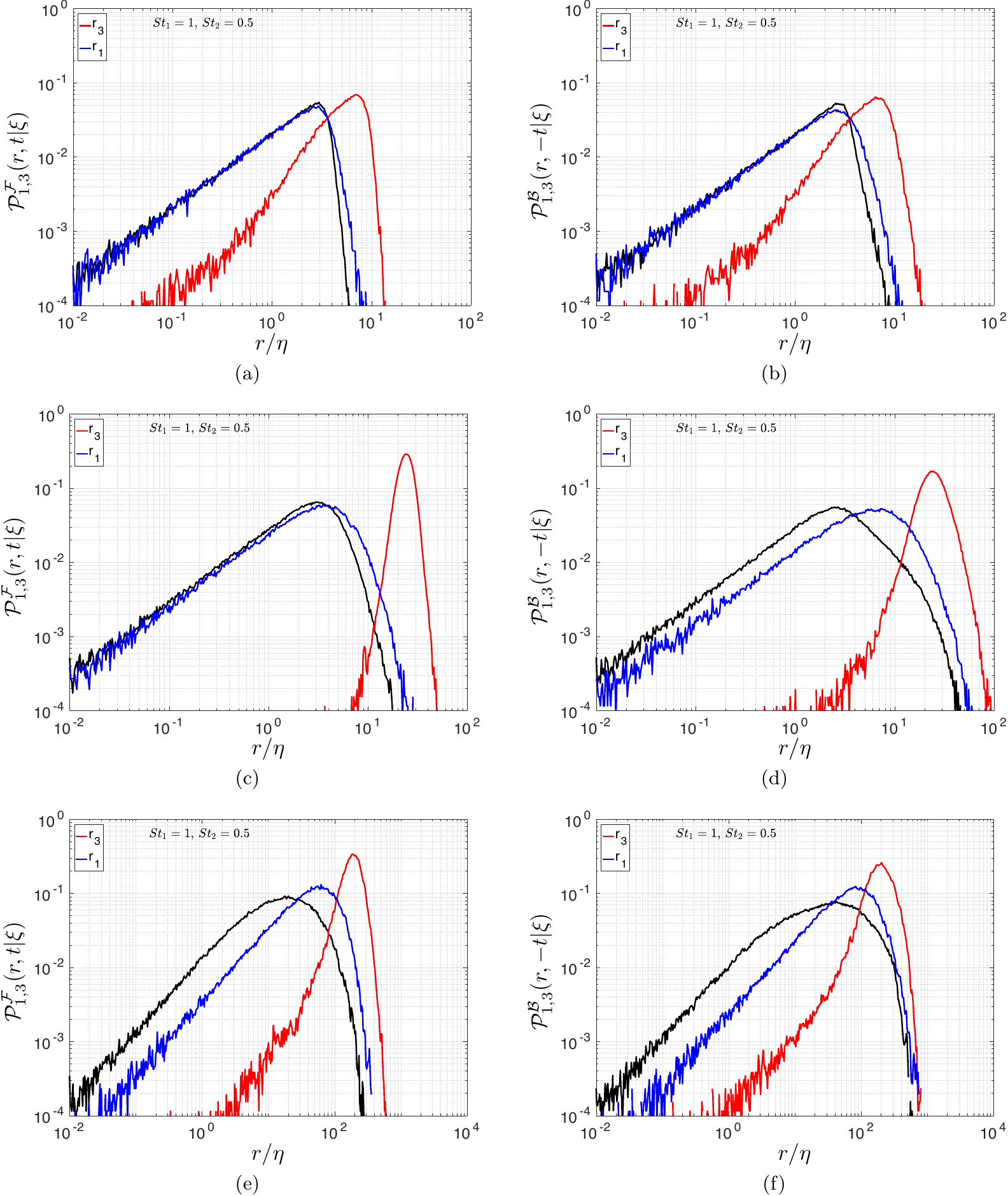}
		\caption{DNS results for $\mathcal{P}^\mathcal{F}_{1,3}(r,t\vert \xi)$ (plots (a,c,e)), $\mathcal{P}^\mathcal{B}_{1,3}(r,-t\vert \xi)$ (plots (b,d,f)) for $St_1 = 1, St_2 = 0.5$ with $\xi \in [3\eta,4\eta]$ and (a,b) $t = 0.5\tau_\eta$, (c,d) $t = 2.5\tau_\eta$, (e,f) $t = 20\tau_\eta$. The red line corresponds to the vertical separations, blue line corresponds to the horizontal separations, and black line corresponds to the results without gravity.}
	\label{fig:pdf1}
	\end{figure}

\begin{figure}
	\centering
	\includegraphics[width=\textwidth]{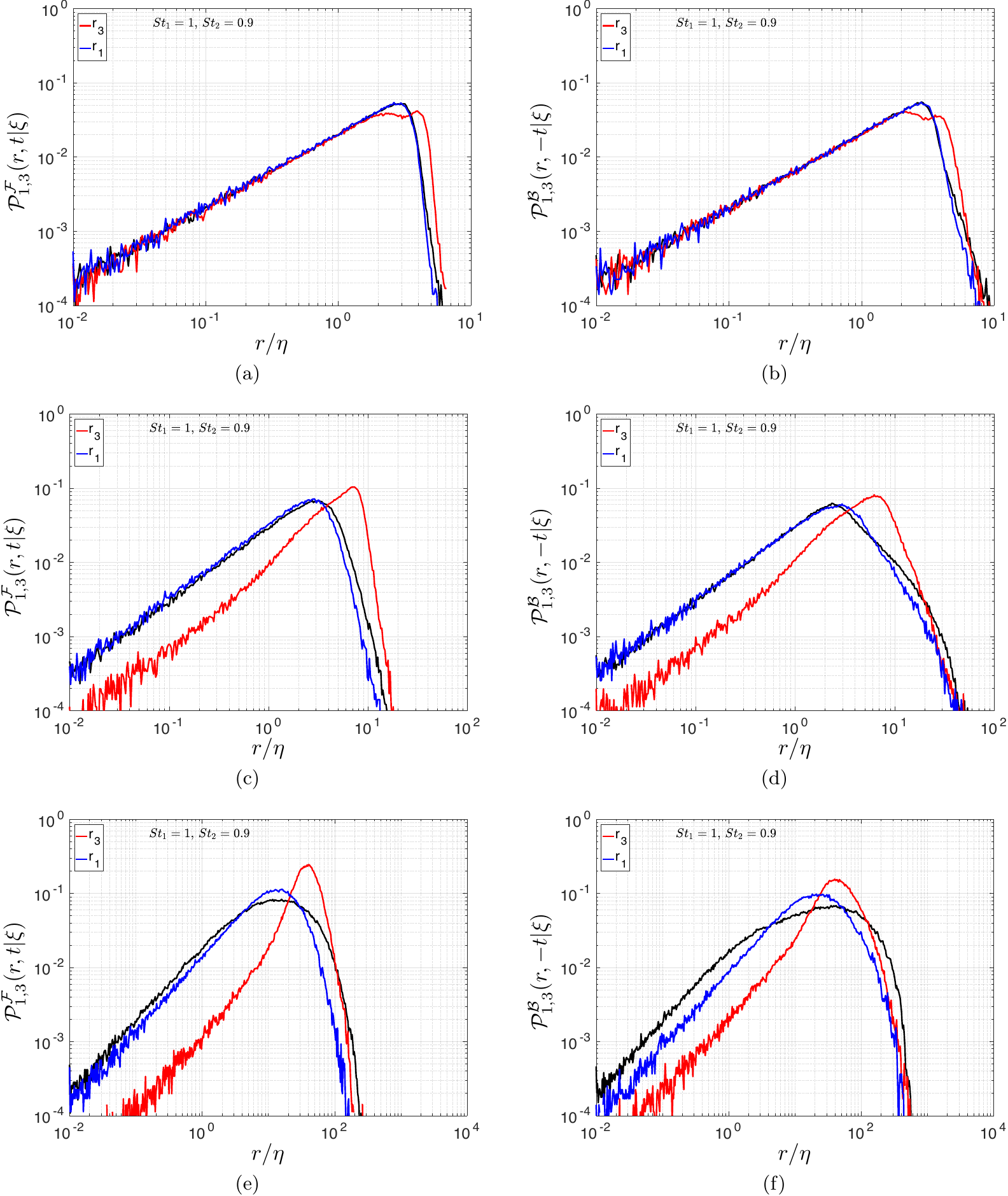}
		\caption{DNS results for $\mathcal{P}^\mathcal{F}_{1,3}(r,t\vert \xi)$ (plots (a,c,e)), $\mathcal{P}^\mathcal{B}_{1,3}(r,-t\vert \xi)$ (plots (b,d,f)) for $St_1 = 1, St_2 = 0.9$ with $\xi \in [3\eta,4\eta]$ and (a,b) $t = 0.5\tau_\eta$, (c,d) $t = 2.5\tau_\eta$, (e,f) $t = 20\tau_\eta$. The red line corresponds to the vertical separations, blue line corresponds to the horizontal separations, and black line corresponds to the results without gravity.}
	\label{fig:pdf2}
	\end{figure}

\begin{figure}
	\centering
	\includegraphics[width=\textwidth]{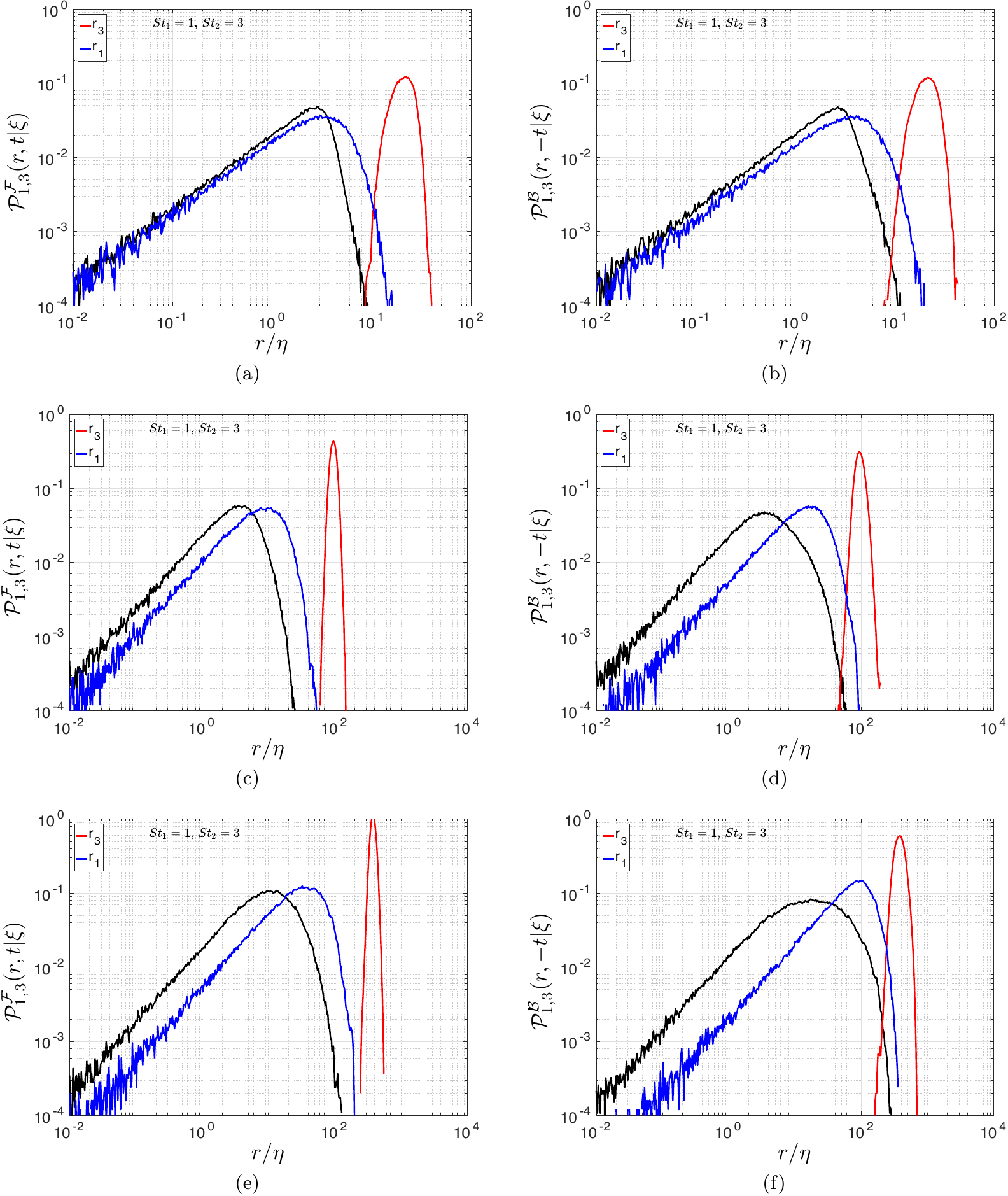}
		\caption{DNS results for $\mathcal{P}^\mathcal{F}_{1,3}(r,t\vert \xi)$(plots (a,c,e)), $\mathcal{P}^\mathcal{B}_{1,3}(r,-t\vert \xi)$ (plots (b,d,f)) for $St_1 = 1, St_2 = 3$ with $\xi \in [3\eta,4\eta]$ and (a,b) $t = 0.5\tau_\eta$, (c,d) $t = 2.5\tau_\eta$, (e,f) $t = 10\tau_\eta$. The red line corresponds to the vertical separations, blue line corresponds to the horizontal separations, and black line corresponds to the results without gravity.}
	\label{fig:pdf3}
	\end{figure}

We now consider the FIT and BIT Probability Density Functions (PDFs) of the horizontal and vertical separations, defined as
\begin{align}
\mathcal{P}^\mathcal{F}_{1}(r,t\vert\xi)&\equiv\langle \delta(|r^p_1(t)|-r) \rangle_\xi,\\
\mathcal{P}^\mathcal{F}_{3}(r,t\vert\xi)&\equiv\langle \delta(|r^p_3(t)|-r) \rangle_\xi,
\end{align}
and similarly for the BIT PDFs $\mathcal{P}^\mathcal{B}_{1}$ and $\mathcal{P}^\mathcal{B}_{3}$. Figures \ref{fig:pdf1}-\ref{fig:pdf3} show the results for these PDFs with initial separation $\xi \in [3\eta,4\eta]$, and for different times. As expected, the results show that gravity affects $\mathcal{P}_{1}^\mathcal{F,B}$ and $\mathcal{P}_{3}^\mathcal{F,B}$ in different ways, since gravity only plays an explicit role in the vertical direction. Consistent with the horizontal and vertical mean-square separations results and the arguments of \S\ref{theory}, gravity shifts the horizontal and vertical separation PDFs towards larger values for pairs with $|\Delta St | \geq 0.5$, and suppresses them for the weakly bidisperse pair ($|\Delta St| = 0.1$) for the value of $\xi$ considered. 

When gravity dominates the vertical dispersion, the following result holds
\begin{align}
\lim_{Fr\to0}\mathcal{P}^\mathcal{F}_3(r,t\vert \xi) &= \delta\Big(|\xi-u_\eta\Delta St Fr^{-1}t|-r\Big),\label{eq:FIT_PDFga} 
\end{align}
and similarly for $\mathcal{P}^\mathcal{B}_3$. Figure \ref{fig:pdf3} shows the results for $|\Delta St| = 2$ and $Fr = 0.052$. According to the non-dimensionalized equation for $\bm{w}^p$, namely \eqref{eq:sol_w}, the differential sedimentation contribution to $\bm{w}^p$ is much larger (an estimate is nearly forty times larger) than the contributions associated with the turbulence when $|\Delta St| = 2$ and $Fr = 0.052$. 
\FloatBarrier
\FloatBarrier
As a result, we might expect that $\mathcal{P}^\mathcal{F}_3(r,t\vert \xi)$ should be close to the delta function prediction in \eqref{eq:FIT_PDFga}. However, the results in figure \ref{fig:pdf3} show that even for this case, $\mathcal{P}^\mathcal{F}_3(r,t\vert \xi)$ shows significant deviations from a delta function. This occurs because although the settling velocity contribution for $|\Delta St| = 2$ and $Fr = 0.052$ is much larger than the ``typical'' velocities of the turbulence, due to intermittency, there are significant regions of the flow where the turbulent velocities are of the same order as the settling velocity. This highlights the limitations of using scaling analysis in turbulence, namely, that because it only considers the mean-field behavior of the system, it cannot faithfully describe how the system behaves during fluctuations of the system about its mean-field behavior. In order to observe the asymptotic behavior of \eqref{eq:FIT_PDFga}, we would need extremely large values of $|\Delta St|/Fr$, and these values would need to be larger as $R_\lambda$ increases due to the increased intermittency of the flow with increasing $R_\lambda$. In most practical applications of particle mixing and transport in turbulence, such values may never be obtained, implying that turbulence will always play an important role in the vertical mixing of settling, bidisperse particles, and its effect cannot be ignored (unless one is only interested in low-order moments of the dispersion process).
\section{Conclusions}\label{conc}
In this paper, we have used DNS to investigate the relative dispersion of settling, bidisperse inertial particles in isotropic turbulence. We also considered differences in the way gravity and turbulence affect the particle relative dispersion in the vertical (parallel to gravity) and horizontal directions. A key motivation for this work stems from the findings of our recent study \citep{dhariwal18jfm}, where we observed a number of non-trivial effects of the combined influence of turbulence and gravity on the vertical and horizontal relative velocities of settling bidisperse particles.

We found that for particles with $|\Delta St| \geq 0.5$, gravity enhances the mean-square separations of the particles both forward in time (FIT) and backward in time (BIT), whereas it suppresses the relative dispersion of weakly bidisperse particles with $|\Delta St| = 0.1$. We also observed that the duration over which the particles separate ballistically is much larger when they are subjected to gravity as compared to the case without gravity, as was also observed in  \cite{chang15} for weakly inertial particles. For Froude number $Fr = 0.052$, the vertical relative dispersion is enhanced for the range of Stokes numbers considered, and the enhancement is primarily due to the differential sedimentation of the particles. On the other hand, gravity has a non-trivial effect on horizontal relative dispersion, enhancing the relative dispersion for particles with $|\Delta St| \geq 0.5$ and suppressing it for weakly bidisperse particle-pairs (i.e., $|\Delta St| \ll1$). This differing behavior arises of the fundamental differences in the mechanisms governing the small-scale relative velocities of bidisperse and monodisperse (and weakly bidisperse) particle-pairs, and how these are affected by gravity. We note, however, that we only considered particles with initial separations $\leq O(10)$ Kolmogorov lengths. For larger initial separations, the range of $|\Delta St|$ for which one would observe either the enhancing or suppressing effect of gravity on the relative dispersion would be different. Nevertheless, the effect would still occur, just in different portions of the parameter space.

Finally, we considered the FIT and BIT PDFs of the horizontal and vertical separations. When $|\Delta St| = 2$ and $Fr = 0.052$, the PDFs of the vertical dispersion show a substantial effect of turbulence on the relative dispersion, despite the fact that the differential sedimentation speed is large. This indicates that even when $|\Delta St| \geq O(1)$ and $Fr \ll 1$, the effect of turbulence on the vertical dispersion cannot be simply based on the differential settling of the particle-pair, as intermittency allows the turbulence to continue to affect the higher-order moments of the statistics characterizing the dispersion process. Indeed, in order to observe the asymptotic behavior of vertical dispersion based purely on the differential settling, one would require extremely large values of $|\Delta St|/Fr$, such that given the parameter regimes in many applications, the effect of turbulence on the vertical mixing of settling, bidisperse particles may never be ignored (unless one is only interested in the low-order statistics of the dispersion process).
\section{Acknowledgements}
This work used the Extreme Science and Engineering Discovery Environment (XSEDE), which is supported by National Science Foundation grant number ACI-1548562 \citep{xsede}. Specifically, the Comet cluster was used under allocation CTS170009.

\bibliographystyle{unsrt}
\bibliography{refs_co12}

\end{document}